\documentclass[]{interact}

\usepackage{epstopdf}
\usepackage[caption=false]{subfig}

\usepackage[numbers,sort&compress]{natbib}
\bibpunct[, ]{[}{]}{,}{n}{,}{,}
\renewcommand\bibfont{\fontsize{10}{12}\selectfont}
\makeatletter
\def\NAT@def@citea{\def\@citea{\NAT@separator}}
\makeatother

\theoremstyle{plain}

\theoremstyle{definition}

\theoremstyle{remark}

\author{
\name{Muhammad~Sajjad Akbar\textsuperscript{a}\thanks{CONTACT Muhammad~Sajjad Akbar. Email: muhammad.akbar@sydney.edu.au} }
\affil{\textsuperscript{a} University of Sydney, Australia; }
}

\begin{document}

\articletype{ARTICLE TEMPLATE}

\title{Beyond Detection: Designing AI-Resilient Assessments with Automated Feedback Tool to Foster Critical Thinking and Originality}

\maketitle

\begin{abstract}

The growing prevalence of generative AI tools such as ChatGPT has raised urgent concerns about their impact on student learning, particularly their potential to erode critical thinking and creativity in academic contexts. As students increasingly use these tools to complete assessments, foundational cognitive skills are at risk of being bypassed, challenging the integrity of higher education and the authenticity of student work. Current AI-generated text detection tools are fundamentally inadequate in addressing this challenge. They produce unreliable, unverifiable outputs and are highly susceptible to false positives and false negatives, especially when students apply obfuscation techniques such as paraphrasing, translation, or structural rewording. These tools rely on shallow statistical features rather than contextual or semantic understanding, making them unsuitable as definitive indicators of AI misuse. In response, this research proposes an AI-resilient, assessment-based solution that shifts focus from reactive detection to proactive assessment design. The solution is delivered through a web-based Python tool that integrates Bloom’s Taxonomy with advanced natural language processing techniques including GPT-3.5 Turbo, BERT-based semantic similarity, and TF-IDF metrics to evaluate the AI-solvability of assignment tasks. By analyzing both surface-level and semantic features, the tool helps educators assess whether a task targets lower-order thinking (e.g., recall, summarization), which is more easily completed by AI, or higher-order skills (e.g., analysis, evaluation, creation), which are more resistant to AI automation. This framework empowers educators to intentionally design cognitively demanding AI-resistant assessments that promote originality, critical thinking, and fairness. By addressing the design of root issue assessment rather than relying on flawed detection tools, this research contributes a sustainable and pedagogically sound strategy to uphold academic standards and foster authentic learning in the era of AI.

\end{abstract}


\section{Introduction}

Integrating AI-technology with innovative thinking skills in higher education (HE) environment has grown more challenging due to rapid digital innovation and ubiquitous data availability. In applied education, innovative thinking is essential. It is characterised by the ability to come up with fresh, original, and practical ideas, methods, and solutions is referred to as innovative thinking. It entails thinking creatively to come up with original solutions to issues, enhance workflows, or open up new possibilities. \cite{curzon2003higher}. 
Fostering critical thinking in higher education is essential for enabling transformative learning. A critical stance involves the ability to reflect, question assumptions, and engage with diverse perspectives. This approach promotes autonomy, self-reflection, and the capacity to challenge existing paradigms.
Barriers to adopting a critical stance include institutional constraints and passive learning habits formed through rote memorization. Effective pedagogical strategies, such as collaborative learning, reflective practices, and dialogical teaching, help address these challenges. Trust between educators and students is critical for fostering open dialogue and meaningful engagement. Developing a critical stance equips students with essential tools for personal growth, independent thought, and active participation in democratic societies.

According to recent research, the emphasis of higher education is shifting from discipline-specific information to the development of creative skills, allaying worries that graduates might not possess these vital competencies \cite{garcia2024soft} .

Soft skills play a central role in graduate education, with creativity, leadership, and analytical thinking frequently emphasized in curricula. An analysis of 230 graduate programs across 49 institutions identified 31 distinct soft skills in learning outcomes, though skills such as empathy, ethical reasoning, and critical thinking were underrepresented.
The variability in the integration of soft skills across disciplines highlights the need for more balanced frameworks. A uniform approach to soft skills development risks neglecting essential competencies specific to different professional contexts. A more comprehensive strategy ensures graduates possess a well-rounded skill set to meet modern industry demands. Holistic integration of soft skills enhances the effectiveness of graduate programs in preparing students for diverse and dynamic professional environments. The AI evolution has a significant impact on the teaching and learning practices for the students and the staff. It can be used to produce a wide range of media types, including text, images, audio, and videos. The technology's adaptability positions it as a critical tool in advancing productivity and creativity. Businesses and institutions are exploring its applications to streamline operations and generate value. Generative AI is rapidly evolving, reshaping the boundaries of what machines can create and augmenting human capabilities in unprecedented ways. Text generators like ChatGPT demonstrate how GAI has the ability to transform education by enhancing learning opportunities and skill development.

\subsection{\textsf{Adapt or Resist}}

The integration of AI text generators like ChatGPT into academic settings has sparked a debate on whether academia should adapt to or resist these technologies. AI text generators offer significant potential to enhance education by automating administrative tasks, assisting with academic writing, and supporting personalized learning. These tools can improve efficiency, reduce workload, and facilitate the creation of accessible learning materials for diverse student populations.
However, challenges such as academic integrity, reliance on AI, and the potential loss of critical thinking skills raise concerns. The technology may enable unethical practices like plagiarism and hinder students' ability to engage deeply with subject matter. Furthermore, biases inherent in AI-generated content and the lack of transparency in algorithms present risks that could compromise the educational process.
To address these issues, academia must strike a balance between adaptation and regulation. Developing ethical frameworks, integrating AI literacy into curricula, and fostering critical awareness are essential steps. By leveraging the strengths of AI while mitigating its risks, institutions can ensure that these tools augment rather than undermine educational goals. The paper argues for a proactive approach, emphasizing the need for a collaborative effort to adapt responsibly to the evolving landscape of AI in education.

This article demonstrates how AI may help students see things from a variety of angles while also encouraging their creativity and critical thinking \cite{bearman2023discourses}. It also examines the diverse discourses surrounding the integration of artificial intelligence (AI) in higher education. AI is presented as a transformative force capable of reshaping teaching, learning, and administration through tools like adaptive learning platforms, automated grading systems, and virtual tutors. These technologies offer opportunities to enhance personalization, improve efficiency, and support evidence-based decision-making in educational practices.
However, the review highlights significant concerns regarding equity, ethics, and the sociocultural implications of AI adoption. It critiques the potential for AI to perpetuate systemic biases, reinforce existing inequalities, and reduce human agency in education. Questions are raised about the over-reliance on data-driven approaches that may neglect the nuances of educational contexts and undervalue critical human interactions.
The paper categorizes the discourses into optimistic, cautious, and critical perspectives. Optimistic views emphasize AI's potential for innovation and progress, while cautious perspectives stress the need for governance, transparency, and ethical safeguards. Critical discourses focus on the risks of commercialization, data privacy concerns, and the commodification of education through AI technologies.
The review concludes that AI in higher education must be approached with critical awareness, balancing its benefits with its risks. Institutions are encouraged to adopt ethical frameworks, involve diverse stakeholders, and prioritize human-centered approaches to ensure that AI technologies support inclusive and meaningful educational outcomes.

But given how quickly new technologies are being included, concerns have been raised regarding how this would affect conventional and traditional learning methods.

Therefore the important question is: Is Artificial Intelligence Really the Next Big Thing in Learning and Teaching in Higher Education?
Artificial intelligence (AI) presents opportunities for transformative change in higher education through intelligent tutoring systems, predictive analytics, and automated feedback mechanisms. These technologies are recognized for their potential to enhance personalized learning experiences, improve scalability, and optimize administrative functions. However, the extent of AI's transformative impact remains uncertain, with claims of widespread innovation often lacking robust empirical evidence \cite{o2023artificial}.
Challenges associated with AI adoption include algorithmic bias, ethical concerns, and an over-reliance on automated systems, which may undermine meaningful human interaction in education. Despite these limitations, AI is increasingly regarded as a complementary tool that can support traditional teaching methods when implemented with care and ongoing evaluation. Effective integration requires alignment with pedagogical goals and equitable access to technology to ensure inclusivity.
The role of AI in education highlights the need for balancing innovation with caution. Ethical frameworks and evidence-based practices are essential to maximize benefits while addressing potential risks. Maintaining the human dimension of teaching and learning is critical to ensuring AI technologies serve as tools for enhancement rather than disruption. This approach emphasizes the importance of thoughtful implementation to achieve sustainable and meaningful outcomes in higher education.

This research aims to investigate how the increasing use of generative AI tools, such as ChatGPT, is impacting student creativity, critical thinking, and academic integrity. In response to the limitations of existing AI-detection technologies which often produce unreliable results and fail to account for obfuscation techniques this study proposes a proactive, design-focused solution. By integrating Bloom’s Taxonomy with advanced natural language processing techniques, including GPT-3.5 Turbo, BERT-based semantic analysis, and TF-IDF metrics, the research introduces a Python-based tool that predicts the AI-solvability of assessment tasks. The goal is to support educators in designing cognitively demanding assessments that are resistant to AI-generated responses. Through this approach, the study seeks to reduce reliance on flawed detection systems, foster deeper learning, and preserve the originality and creative engagement essential to higher education.

We address the following research questions in order to close this gap:

1. How are postgraduate students’ critical thinking skills influenced by AI text generators at the University of Sydney?

2. What are the limitations of current AI-generated text detection tools in accurately identifying AI-authored academic work, particularly when obfuscation techniques are used?

3. How can Bloom’s Taxonomy be applied to assess the cognitive complexity of assessment tasks in order to determine their vulnerability to AI-generated responses?

4. To what extent can a Python-based solution help educators predict and minimize AI-solvability in assessments through automated cognitive analysis?

5. How does redesigning assessments based on cognitive complexity improve academic integrity and reduce reliance on unreliable AI detection tools?

The following section presents a review of related work on generative AI tools and the associated challenges they pose in educational contexts. This is followed by the Problem Identification section, which provides an evidence-based examination of how such tools are impacting student creativity. The subsequent sections outline the proposed solution and detail its technical implementation. Finally, the Results are discussed in the Analysis and Comparison sections, highlighting the effectiveness of the solution in addressing the identified issues.
\section{Related Work}


This study examines the impact of Generative AI (GAI), specifically ChatGPT, on critical thinking skills among UK postgraduate business school students using Bloom’s taxonomy as a framework. A mixed-method approach with 107 participants reveals that AI enhances lower-order cognitive skills (e.g., remembering and understanding) but raises concerns about reliability, accuracy, and ethics. While the study offers valuable insights into pedagogy, policy, and practice, its limited sample size restricts generalizability, and the UK-centric approach overlooks global socio-economic factors. Additionally, relying solely on Bloom’s taxonomy may not fully capture the diverse aspects of critical thinking, and ethical concerns such as accessibility and the digital divide remain underexplored. Future research should include larger, more diverse samples, adopt broader competency frameworks, and analyze global variations in AI’s impact on education to ensure a comprehensive understanding of AI’s role in fostering critical thinking skills \cite{essien2024influence}.

Artificial Intelligence (AI) has significantly transformed educational assessment practices, particularly through AI-based assessment generators, which enhance efficiency, accuracy, and scalability in creating and grading assessments. This study explores the integration of AI in educational assessments, highlighting its ability to provide impartial feedback, improve learning outcomes, and streamline evaluation processes. However, several critical challenges must be addressed, including biases in AI models that may reinforce systemic inequities, a lack of transparency in AI decision-making that raises accountability concerns, and ethical issues related to data privacy and security. Additionally, an over-reliance on AI-generated assessments risks diminishing pedagogical integrity by reducing human oversight in evaluating student performance. To fully harness the benefits of AI in education, it is crucial to implement safeguards that promote fairness, inclusivity, and explainability while ensuring AI-driven assessments align with educational values. Future research should focus on refining AI algorithms to mitigate biases, improve transparency, and enhance adaptability, enabling AI-based assessment generators to create more equitable and effective evaluation frameworks \cite{boutyour2024artificial}.

ChatGPT, as the most advanced chatbot to date, has sparked both excitement and concern regarding its impact on higher education, particularly in assessment, learning, and teaching. Developed by OpenAI as part of the Generative Pretrained Transformer (GPT) language model, ChatGPT is capable of generating human-like text and engaging in natural conversations. This article explores the evolution of OpenAI, highlighting its transition from a non-profit entity to a commercial enterprise. Through an extensive literature review and experimental analysis, the study examines ChatGPT’s strengths, limitations, and its implications for higher education. While AI-driven chatbots offer promising applications for students, educators, and academic systems, they also pose challenges such as ethical concerns, academic integrity issues, and potential disruptions to traditional assessment methods. The study positions ChatGPT within the broader field of Artificial Intelligence in Education (AIEd), assessing both opportunities and threats, and concludes with recommendations for students, teachers, and institutions to adapt to the evolving educational landscape shaped by AI chatbots \cite{rudolph2023chatgpt}.

The emergence of ChatGPT 3.0 in November 2022 has sparked widespread debate on AI’s role in student learning and assessment, with models like Llama, Gemini, and CoPilot being increasingly used for generating concise responses from vast datasets. Over the past 18 months, AI capabilities have expanded beyond text synthesis to include image, voice, and video generation, raising concerns about the authenticity of student work and the rise of AI-assisted plagiarism. Studies show that over 50\% of students have considered using AI for academic purposes, though most disapprove of AI-generated assignments. In response, the Australian Tertiary Education Quality and Standards Agency (TEQSA) called for a reassessment of teaching and assessment practices, prompting a national inquiry and initial AI bans in Australian universities, including The University of Sydney However, recognizing AI’s potential, the Educational Innovation Portfolio encouraged controlled AI trials within assessments \cite{charles2025ai}.

The rapid growth of various types of data has opened new possibilities for analyzing educational trends, particularly in assessing how AI-generated text influences student performance. This research investigates the impact of assessment on student knowledge by analyzing a dataset collected from open-question responses during a midterm exam at Vilnius Gediminas Technical University (VILNIUS TECH). Statistical analysis revealed that students who utilized text generators achieved higher grades by paraphrasing AI-generated answers effectively. The study also identified question types that students struggled with when answering independently, highlighting the potential for AI to assist in complex problem-solving. As interest in Educational Data Mining (EDM) and Learning Analytics (LA) grows, the availability of generative AI, such as ChatGPT, has altered assessment practices, raising concerns about academic integrity and fair grading. This research contributes to understanding how AI-generated responses influence student performance, comparing paraphrased AI-assisted answers with original responses. The findings indicate that allowing students to use text generation tools significantly increases the percentage of top grades, demonstrating AI’s impact on assessment outcomes. However, further investigation is needed to determine factors that contribute to lower AI-generated answer accuracy, as only one question had an average score below 75\%. These insights can help educators refine assessment tasks and establish guidelines for ethical AI integration in student evaluations \cite{pliuskuviene2024educational}.

Artificial Intelligence (AI) tools have become a significant topic of debate in higher education, with some faculty integrating them into coursework while others oppose their use in academic assignments. Despite varying institutional stances, students are actively utilizing AI, raising concerns about academic integrity and the need for clear guidelines on its ethical application. This study explores faculty perspectives on AI in undergraduate literature, language, and linguistics programs, revealing key themes such as AI implementation, integrity concerns, and its implications for the humanities. While AI has the potential to enhance writing and learning, professors emphasize the importance of responsible and strategic use, ensuring that students develop essential linguistic skills rather than relying solely on AI-generated content. However, limitations such as time constraints, sample size, and media bias influenced this research. Future studies should incorporate perspectives from a broader range of disciplines and students to assess the evolving role of AI in education while ensuring a balanced approach between technology and human creativity \cite{yitages2024faculty}.

Large language models (LLMs) enable students to generate seemingly well-crafted assessment responses, raising concerns about their impact on academic integrity. Despite limitations, LLM detection tools offer insights into the extent of AI usage in university assessments. This study analyzes 10,725 student assessments from two cohorts (2022 and 2023) using the TurnItIn AI detection tool, revealing an increase in flagged submissions, suggesting the growing use of generative AI in academic writing. Additionally, demographic analysis indicates that male students, younger students, and those with lower prior educational attainment are more likely to have their work flagged as AI-generated. While detection rates increased, they remained slightly lower than TurnItIn’s reported 20\% benchmark. These findings provide a baseline for understanding AI adoption trends, offering valuable insights for refining assessment strategies and AI detection methods as generative AI becomes more prevalent in student submissions. The study underscores the need for continued monitoring and adaptation of academic policies in response to evolving AI use \cite{gooch2024exploring}.

Ethical frameworks for text generators (TGs) in education typically address concerns such as personalized learning, data dependency, biases, academic integrity, and the potential reduction of student creativity. While institutional guidelines provide a broad ethical perspective on AI use, there is a need for a localized approach that considers the interactions between teachers and students. This study explores how an ethical framework for AI use emerges within individual classrooms through a qualitative case study of a composition student’s engagement with ChatGPT as a writing partner. Findings reveal the concept of a “local ethic,” a dynamic, student-teacher negotiated framework that shapes AI’s role in learning environments. The study recommends that educators foster critical discussions on AI authorship, agency, and reliability while adapting assignments and policies to reflect these evolving ethical considerations \cite{vetter2024towards}.

The emergence of ChatGPT has sparked significant concern and debate in higher education regarding its impact on student assessments and academic integrity. As a tool capable of generating human-like written content, ChatGPT has gained widespread attention from students, leading to fears of misuse in formative and summative assessments. This study reviews how and why students use ChatGPT to support academic work, explores challenges in detecting AI-generated content, and critically considers the ethical implications and pedagogical opportunities AI presents. While educators and policymakers acknowledge ChatGPT's potential to improve academic efficiency such as summarizing text, drafting abstracts, and solving complex queries concerns remain about its role in undermining authentic student performance, quality assurance, and graduate employability. The research highlights the need for collaborative efforts across educational institutions to develop ethical guidelines, foster AI literacy, and rethink assessment strategies to include higher-order thinking tasks. Importantly, it calls for more research into students' experiences and motivations, emphasizing the need for informed, responsible use of AI in academic settings \cite{gamage2023chatgpt}.

The reviewed literature collectively highlights a complex relationship between AI technologies and higher education. While generative AI tools such as ChatGPT demonstrate potential in enhancing lower-order cognitive tasks and improving assessment efficiency, they also introduce substantial ethical, pedagogical, and policy challenges. A recurring limitation across existing studies is the over-reliance on detection-based strategies and the narrow application of frameworks like Bloom’s Taxonomy. Most research emphasizes the need for broader, context-aware approaches that incorporate classroom-level ethics, diverse assessment design, and AI literacy. These findings justify the need for a proactive, design-oriented solution such as the one proposed in this study that integrates cognitive complexity analysis, semantic evaluation, and real-time solvability prediction to ensure academic integrity and promote authentic learning experiences in AI-integrated educational environments.

\section{Evidance: AI tools Affecting Cognitive Skills of the Students}

Bloom's taxonomy is a useful framework for assessing educational learning objectives in order to comprehend how AI text generators can improve critical thinking abilities \cite{banda2023application}. It reflects the incremental complexity inherent in critical thinking by classifying cognitive processes into discrete levels, including remembering, comprehending, applying, analysing, evaluating, and producing.

The use of AI text generators in the assessments has brought out particular difficulties. Empirical studies examining how AI text generators affect critical thinking abilities are scarce, especially when it comes to postgraduate education. This offers a strong case for more investigation and to incorporate that we conducted an in-class anonymous survey at University of Sydney after the submission date of the first individual assignment of a computer science unit. Participants were enquired about the effect of using GAI tools (ChatGPT etc.) on the learning process. The assessment was designed in a way that requires innovative capabilities. We used a single survey instrument to capture both qualitative and quantitative data in order to have a thorough grasp of the participants' experiences and perceptions about employing AI in learning.

\subsection{\textsf{Survey Questions}}
The Bloom's Taxonomy framework, a commonly accepted taxonomy in the field of educational psychology for assessing cognitive capacities, served as the foundation for the quantitative evaluations in our data collection instrument.
Survey has the following questions based on Bloom taxonomy framework: 

Question 1: Understanding: How do you think the use of AI text generators influenced your understanding of the subject matter in your assessment?

Question 2: Applying: Describe a specific instance where you incorporated an idea generated by an AI text tool into your assessment. How did you adapt or modify the AI's output to fit your own writing style and the requirements of the assignment?

Question 3: Analyzing: Compare and contrast an assessment you completed with the help of an AI text generator and one you completed without it. What differences do you notice in terms of originality, depth of thought, and overall creativity?

Question 4: Evaluating: In your opinion, does the use of AI text generators enhance or hinder your creative process? Provide reasons and examples to support your answer.

Question 5: Creating: Imagine you are tasked with creating an original piece of work without the aid of AI text generators. How would you approach this task differently compared to when you used AI assistance? What steps would you take to ensure that your creativity is fully expressed?

\subsection{\textsf{Analysis and validation of the Problem}}
 
Thematic Analysis of Survey Responses
Thematic analysis was conducted using Braun and Clarke’s six-phase approach to analyze students' responses on the use of AI text generators in their assessments. This framework helped identify key themes and categories related to the impact of AI on understanding, application, analysis, evaluation, and creativity.

\begin{table}[hbt!]
\begin{tabular}{|l|l|l|}
\hline
\textbf{Category} & \textbf{Theme}                                                                & \textbf{Description}                                                                                                                                                                                                                    \\ \hline
Understanding     & Dependency on AI                                                              & \begin{tabular}[c]{@{}l@{}}Many students reported that AI \\ improved understanding \\ but led to reliance on AI-generated \\ explanations rather$\sim$than independent \\ analysis.\end{tabular}                                       \\ \hline
                  & \begin{tabular}[c]{@{}l@{}}Independent \\ Comprehension\end{tabular}          & \begin{tabular}[c]{@{}l@{}}Some students felt they understood \\ the content independently, \\ reflecting self-directed learning without\\ $\sim$the need for AI assistance.\end{tabular}                                               \\ \hline
Applying          & Minimal Adaptation                                                            & \begin{tabular}[c]{@{}l@{}}Students often reported only minor \\ modifications to AI outputs, \\ indicating limited personal input and \\ reliance on pre-existing \\ structures from the AI.\end{tabular}                              \\ \hline
                  & \begin{tabular}[c]{@{}l@{}}Collaborative \\ Refinement\end{tabular}           & \begin{tabular}[c]{@{}l@{}}Some students used AI to refine their ideas, \\ showing a balance \\ between AI assistance and their own input\\ $\sim$to enhance personal writing style.\end{tabular}                                       \\ \hline
Analyzing         & \begin{tabular}[c]{@{}l@{}}Perceived Accuracy \\ of AI\end{tabular}           & \begin{tabular}[c]{@{}l@{}}A majority of students believed AI-generated \\ answers were more$\sim$accurate than their own,\\ $\sim$which could reduce confidence in their \\ ability to generate accurate responses.\end{tabular}       \\ \hline
                  & \begin{tabular}[c]{@{}l@{}}Trust in Self-generated \\ Content\end{tabular}    & \begin{tabular}[c]{@{}l@{}}A minority felt that their own responses were \\ more accurate,$\sim$$\sim$suggesting a belief in the value\\ $\sim$of personal thought processes \\ and originality over AI-generated content.\end{tabular} \\ \hline
Evaluating        & \begin{tabular}[c]{@{}l@{}}Creativity Inhibited \\ by AI\end{tabular}         & \begin{tabular}[c]{@{}l@{}}Most students felt AI hindered their creativity,\\ $\sim$leading them to rely \\ on AI-generated ideas instead of original ones.\end{tabular}                                                                \\ \hline
                  & \begin{tabular}[c]{@{}l@{}}Enhanced Creativity\\  for Some\end{tabular}       & \begin{tabular}[c]{@{}l@{}}A few students noted that AI sparked creativity, \\ helping them explore new perspectives \\ and ideas they$\sim$had not previously considered.\end{tabular}                                                 \\ \hline
Creating          & \begin{tabular}[c]{@{}l@{}}Preference for \\ Systematic Research\end{tabular} & \begin{tabular}[c]{@{}l@{}}Without AI, students indicated they would prefer\\ $\sim$in-depth research,$\sim$showing a preference for \\ creativity $\sim$and originality when not dependent \\ on AI assistance.\end{tabular}           \\ \hline
                  & \begin{tabular}[c]{@{}l@{}}Shortcuts over \\ Originality\end{tabular}         & \begin{tabular}[c]{@{}l@{}}Students using AI frequently relied on \\ it as a shortcut for idea generation,$\sim$which\\  may deter them from exploring \\ their creative potential.\end{tabular}                                        \\ \hline
\end{tabular}
\end{table}

\subsection{Impact of AI on Understanding}

Figure 1 shows that 70 percent of students felt that AI tools helped their understanding, while 30 percent believed they comprehended the subject independently. The significant portion of students who relied on AI to enhance understanding suggests that these tools can be beneficial in clarifying concepts, but it also hints at a potential over-reliance on AI-generated explanations. This dependency might limit students' critical thinking and problem-solving abilities in the long term, as they may increasingly depend on AI to interpret content rather than engaging deeply with the material themselves.
 
In contrast, the 30 percent who maintained independent understanding reflects a segment of students who prioritize self-reliance, which is critical for developing autonomous learning skills. This data emphasizes the need to balance AI-assisted learning with strategies that foster independent comprehension, encouraging students to rely on their abilities without always turning to AI for answers.

\begin{figure}[hbt!]
\centering
\subfloat[Impact on Understanding]{%
\resizebox*{7cm}{!}{\includegraphics{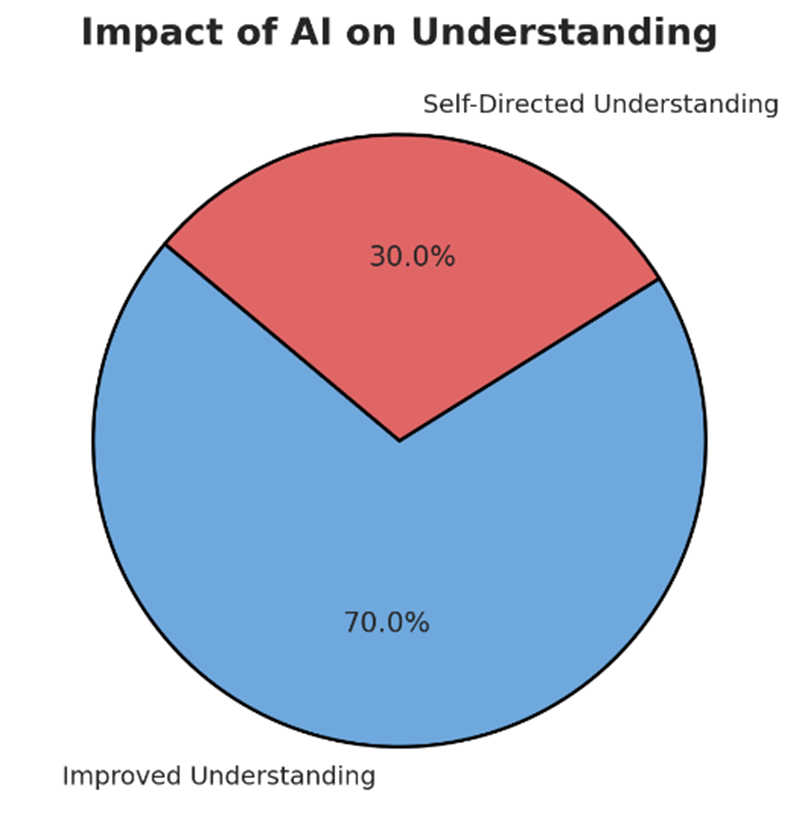}}}\hspace{5pt}
\subfloat[Degree of Modification in using AI Outputs]{%
\resizebox*{8cm}{!}{\includegraphics{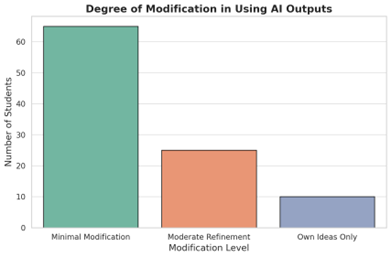}}}
\caption{Impact of AI tools on Understanding and Modification} \label{sample-figure}
\end{figure}

\subsection{Degree of Modification in Using AI Outputs}

Figure 1 shows the extent of students’ modifications to AI-generated content. 65 percent of students reported only minimal changes, suggesting that they may rely heavily on AI-generated text with limited personal input, which could impede the development of unique perspectives and critical thinking skills. When students modify content only slightly, they are less likely to engage in deeper analysis or add personal insights, thus limiting creativity and intellectual growth.
 
Conversely, 25 percent of students used AI to refine their ideas, indicating a more balanced approach where AI is a support tool rather than the sole source of content. These students benefit from the AI’s suggestions but maintain control over their expression, which can enhance creativity by providing a collaborative tool rather than a replacement for personal effort.
The 10 percent who used their own ideas exclusively show that some students still prioritize original thought over AI-generated solutions. This approach is likely to encourage deeper engagement with the material, fostering independent learning and innovation.

\subsection{Perceived Accuracy of AI vs. Own Answers}

Figure 2 reveals that 75 percent of students found AI-generated answers more accurate than their own, while 25 percent believed their responses were more accurate. The high reliance on AI for accuracy could lead students to undervalue their analytical abilities, potentially reducing confidence in their work. When students assume AI’s responses are inherently more accurate, they may feel less motivated to explore and defend their unique perspectives.
The 25 percent who trusted their own answers are indicative of students who value personal insight and critical analysis over AI outputs, which is essential for building self-confidence and critical engagement. This group represents students who may be better equipped to develop original ideas, as they are willing to rely on their understanding rather than defaulting to AI as the “better” option.
 
This graph highlights a potential issue: while AI tools can offer factual accuracy, they should not replace students’ efforts to critically evaluate and trust their own knowledge. Educators may need to encourage students to validate AI-generated content rather than viewing it as an automatic authority.

\begin{figure}
\centering
\subfloat[Impact on the Creativity]{%
\resizebox*{7cm}{!}{\includegraphics{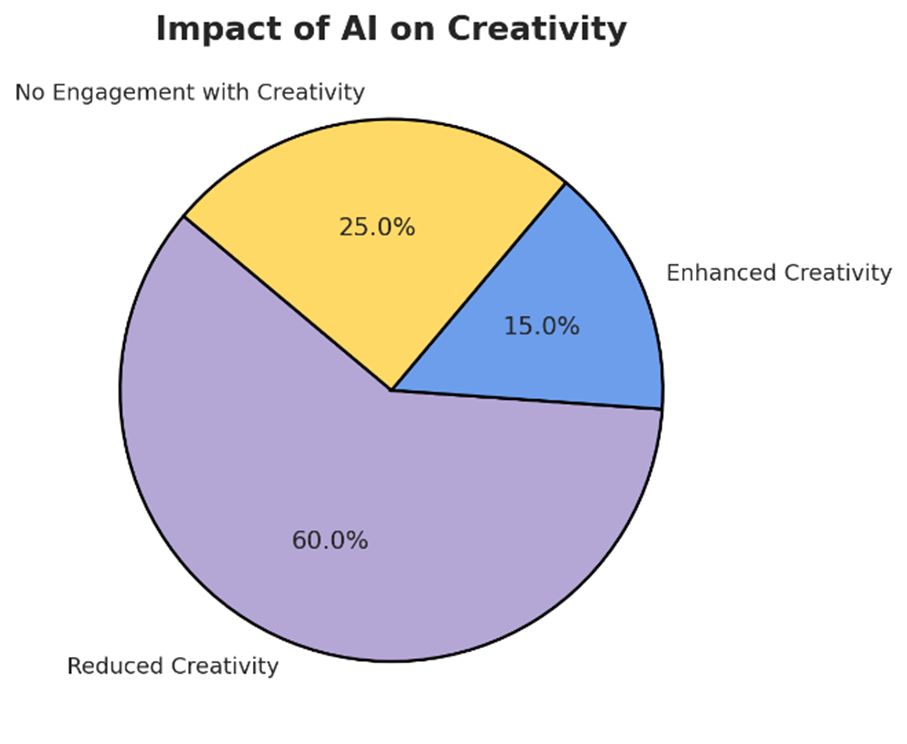}}}\hspace{5pt}
\subfloat[Perceived Accuracy]{%
\resizebox*{7cm}{!}{\includegraphics{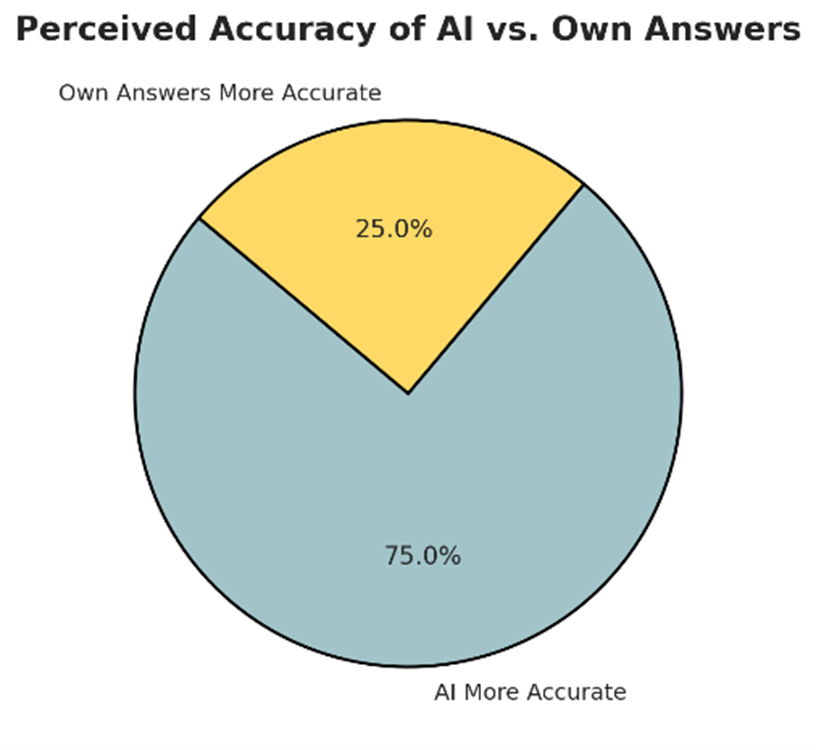}}}
\caption{Impact of Creativity and Perceived Accuracy} \label{sample-figure}
\end{figure}

\subsection{ Impact of AI on Creativity}

In Figure 2, 60 percent of students reported that AI hindered their creativity, 15 percent felt that AI enhanced it, and 15 percent had a neutral view. This data suggests that for most students, AI text tools may limit the creative process, as they rely on AI-generated ideas rather than brainstorming their own. This over-reliance on AI could lead to a loss of originality, as students may bypass the steps of ideation and innovation, simply rephrasing or slightly modifying AI-generated responses.
The 15 percent who found AI beneficial for creativity likely used the tool as a brainstorming assistant, where AI provides prompts or alternative ideas that students then build upon. This use of AI is more collaborative and supports the notion that, when used appropriately, AI can serve as a creative aid rather than a replacement for independent thinking.

The 15 percent with neutral views reflect those who might use AI occasionally but do not see it as significantly impacting their creativity. This balance indicates that while AI can be a helpful tool, its use must be monitored to ensure it supports rather than stifles students’ creative growth.

\subsection{Overall Significance}
Each graph provides insight into the nuanced ways students interact with AI text tools:
\begin{itemize}
  \item Understanding: The first chart suggests that while AI tools can improve comprehension, they risk creating dependency. Educators should emphasize a balanced approach that encourages independent thought. \verb|\item| command.
  \item Application: The second chart indicates a potential lack of engagement with content when students make minimal modifications to AI output. Encouraging students to refine AI ideas further could foster deeper critical engagement.
  \item Accuracy: The third chart warns of potential over-reliance on AI for accuracy, which may affect students' confidence in their work. Developing validation skills can help students trust their knowledge and analytic abilities.
  \item Creativity: The fourth chart highlights the risk of AI stifling creativity for a majority of students. This calls for strategies that promote AI as a supportive tool rather than a primary source of ideas.
\end{itemize}
	
In summary, these results reveal that AI text tools, while helpful for efficiency and quick comprehension, can also hinder students' creative and critical thinking skills if overused. Educators should guide students on how to use AI responsibly, ensuring that it supplements rather than substitutes their creative and analytical processes.

\section{Proposed Solution}

Amid rising concerns over the use of AI-generated content in education, this study assesses the reliability of detection tools designed to differentiate between human and AI-authored text. Analyzing 12 open-access tools and two commercial systems, as shown by the Figure 3, including Turnitin, the findings reveal that most tools are inaccurate and biased toward falsely classifying AI text as human-written. The problem worsens when obfuscation or translation techniques are applied. Despite growing interest in policing AI use, the study concludes that current detection tools are insufficient and potentially misleading in academic contexts.

\begin{figure}[hbt!]
\centering
\subfloat[Accuracy of AI Tools \cite{weber2023testing}]{%
\resizebox*{14cm}{!}{\includegraphics{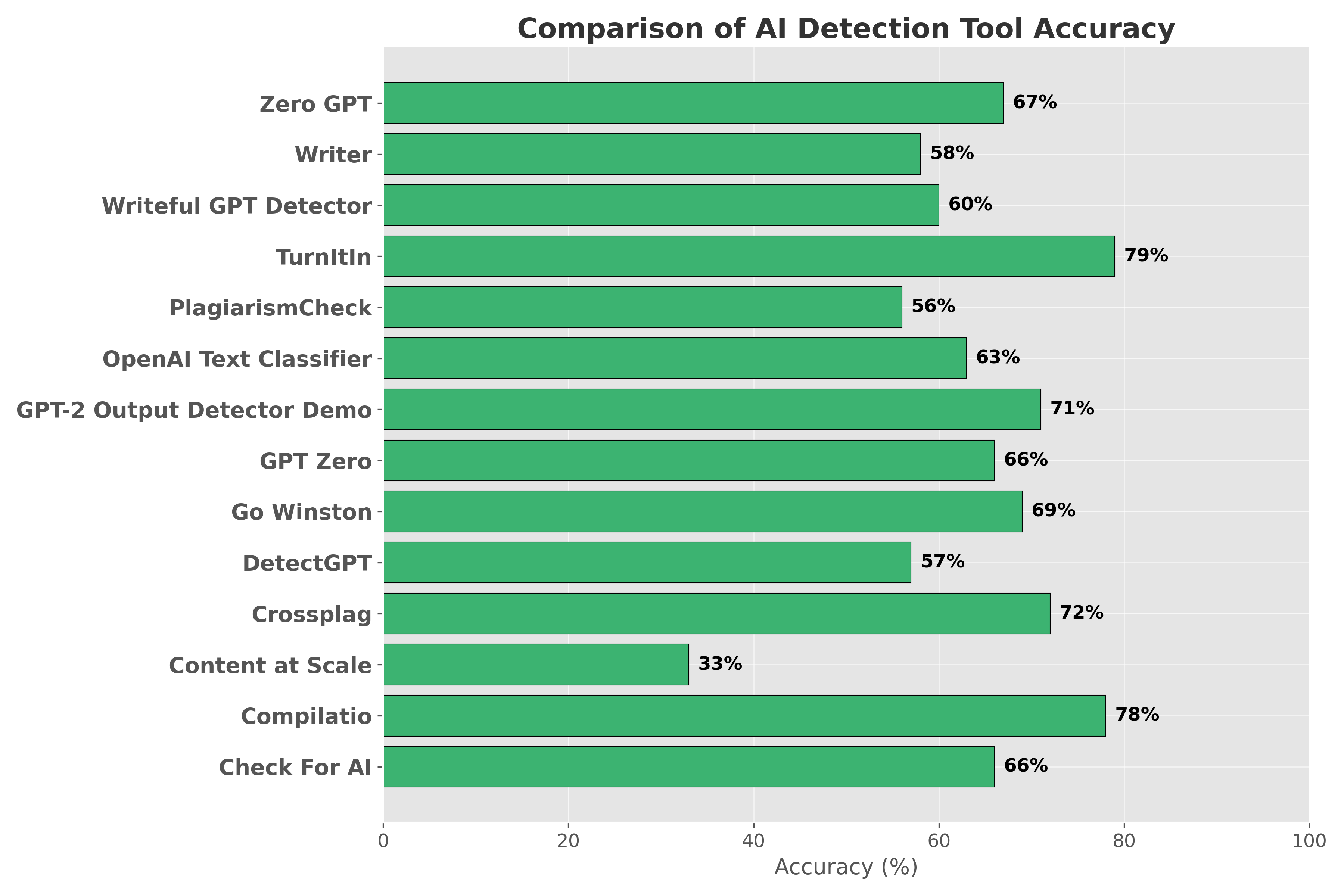}}}\hspace{8pt}
\caption{Accuracy of AI Tools} 
\end{figure} 

AI-generated text detection tools are generally unreliable, with none achieving more than 80\% accuracy and only five surpassing the 70\% threshold. These tools frequently produce false positives misclassifying human-written texts as AI-generated and false negatives, where AI-generated content is incorrectly identified as human-written. Such inconsistencies raise serious concerns about their validity in academic contexts. The findings align with earlier research \cite{gao2022comparing} \cite{anderson2023ai} \cite{elkhatat2023evaluating} \cite{demers16best} \cite{gewirtz2023can} \cite{krishna2023paraphrasing} \cite{pegoraro2023chatgpt}, and sharply contrast with the accuracy claims made by commercial detection tools \cite{walters2023effectiveness} \cite{vsigutevaluation} \cite{weber2023testing} \cite{pegoraro2023chatgpt}. A notable bias is observed across these systems, as they tend to label AI-generated text as human-written, with approximately 20\% of AI content likely being misattributed to human authorship.

These tools lack robustness, as their accuracy significantly declines when faced with obfuscation methods such as manual rewriting or machine paraphrasing. Additionally, they are ineffective at handling texts that have been translated from other languages. In fact, nearly 50\% of AI-generated texts that have been modified through such techniques are likely to be incorrectly identified as human-written. Moreover, the outputs from these detection tools are often difficult for the average user to interpret. While some offer statistical metrics to support their classification, others simply highlight sections of text deemed “likely” to be machine-generated, without providing clear or actionable explanations.

AI-generated text detection tools typically produce vague and unverifiable outputs, such as “This document was likely written by AI” or “11\% likelihood this content originated from GPT-3, GPT-4, or ChatGPT.” These statements lack transparency and do not provide concrete evidence to support their claims. As a result, if a student is accused of using AI-generated content based solely on these outputs, they would have no clear means of defending themselves. The tools demonstrate a wide range of error rates, with false positives ranging from 0\% (Turnitin) to 50\% (GPT Zero), and false negatives from 8\% (GPT Zero) to 100\% (Content at Scale). These failures carry serious consequences false positives risk unjustly penalizing students, while false negatives enable unauthorized use of AI tools, giving some students an unfair advantage and encouraging academic dishonesty. Despite these issues, many educators place unwarranted trust in these tools, but the findings clearly indicate that detection results must be interpreted with great caution. Since AI detection tools do not offer concrete evidence, the ability of educational institutions to conclusively prove academic misconduct based on these tools alone is extremely limited. The reports generated by such tools should not be treated as definitive proof of cheating. At best, they may serve as an initial indication that warrants further attention. Therefore, any suspicion raised by AI detection tools should be followed by meaningful dialogue and discussion with the student, rather than used as the sole basis for disciplinary action. This approach promotes fairness and ensures that students are given the opportunity to explain or clarify their work.

Given the significant limitations of existing AI-generated text detection tools, our decision to introduce a web-based Python solution rooted in Bloom’s Taxonomy emerges as a proactive and pedagogically sound alternative. Current detection technologies face critical challenges: they demonstrate low reliability, lack transparency, and are highly susceptible to obfuscation techniques such as paraphrasing or translation. Most tools fail to surpass even 80\% accuracy, with some misclassifying up to 50\% of AI-generated content as human-written. Their outputs are often vague, unverifiable, and devoid of actionable evidence, making them inadequate as the sole basis for academic misconduct decisions. Furthermore, these tools typically rely on shallow statistical heuristics rather than deep contextual understanding, producing results that are difficult to interpret and easy to misapply.

Instead of focusing on detecting AI use after submission a process fraught with ethical and technical shortcomings our approach tackles the issue at its root: the design of the assessment tasks themselves. We argue that the most sustainable and effective way to uphold academic integrity in the era of generative AI is to create assessments that are inherently resistant to being solved by AI tools.

Our proposed solution not only leverages Bloom’s Taxonomy to analyze the cognitive complexity of assessment tasks but also integrates modern NLP techniques \cite{weber2023testing} \cite{gagliardi2024natural} \cite{carrasco2024enhancing} to evaluate AI-solvability more comprehensively. Specifically, it combines OpenAI’s GPT-3.5 Turbo \cite{gue2024evaluating} \cite{wang2024investigation}, BERT-based semantic similarity analysis \cite{zunlan2025dynamic} \cite{chen2025semantic}, and TF-IDF linguistic complexity metrics \cite{lu2024enhancing} \cite{al2024comparative} to determine how easily a given task could be completed by AI. By incorporating these advanced models, the system can assess not only the surface-level features of a question but also its semantic depth, linguistic difficulty, and conceptual originality. As a result, it calculates an AI-solvability percentage that helps educators identify whether a task promotes lower-order thinking (e.g., recall, summarization) or higher-order thinking (e.g., analysis, evaluation, creation) the latter being much more challenging for AI to replicate.

The effectiveness of our AI-solvability detection framework lies in its ability to merge pedagogical theory with automation efficiency and robust natural language understanding. By doing so, it empowers educators to design meaningful and AI-resilient assessments that prioritize critical thinking, synthesis, and original expression. This shift from reactive detection to proactive design not only circumvents the pitfalls of false accusations and student distrust but also enhances the overall learning experience by aligning assessment practices with educational best standards.

Furthermore, the tool enhances efficiency and scalability in academic settings by automating the assessment process through Python. Using PyMuPDF, the system extracts assignment content from PDF files with high precision. The extracted text is analyzed through GPT-3.5 Turbo to evaluate question difficulty and solvability. Regex-based pattern matching is then employed to extract the AI-solvability percentage from model outputs, while TF-IDF and BERT jointly support the ranking of tasks in terms of AI susceptibility. This end-to-end automation not only reduces the burden of manual review but also empowers instructors to receive real-time, actionable feedback on the design of their assessments. Ultimately, the solution offers a scalable, pedagogically sound approach for strengthening academic integrity in an era increasingly shaped by generative AI.

Ultimately, rather than relying on flawed and potentially unfair AI detection mechanisms, we advocate for equipping educators with intelligent tools that help them construct assessments that generative AI cannot easily complete. This proactive approach reduces opportunities for misuse, reinforces academic integrity, and strengthens the quality and authenticity of learning in an AI-enhanced educational landscape.

\section{Analysis and Comparison}

\subsection{Validation using Assessments}

To validate the effectiveness and applicability of the proposed AI-solvability detection framework, we adopted a data-driven qualitative methodology grounded in real-world academic assessments. A total of 50 assignment tasks were collected from a range of undergraduate and postgraduate units offered by the School of Computer Science at the University of Sydney. These units span multiple subdomains within computer science, including Information Technology, Cybersecurity, Data Privacy, Software Development, and Project Management. The assignments were selected to ensure diversity in content, structure, and cognitive demands, thereby providing a representative sample for testing the scalability and generalizability of our tool across academic contexts.

Each assignment was processed through our web-based Python solution, which extracts textual content from the assignment PDFs, analyzes the complexity of questions using Bloom’s Taxonomy, and generates an AI-solvability score using a combination of GPT-3.5 Turbo, BERT-based semantic similarity, and TF-IDF metrics. The tool categorizes assignments according to their predicted vulnerability to AI-generated responses and offers feedback on the presence of lower-order (e.g., remembering, understanding) versus higher-order (e.g., evaluating, creating) cognitive tasks.

The distribution of AI-solvability scores across the 50 collected assignments is visualized in Figure 4. As shown, the majority of assignments (22) fall within the 65–74\% AI-solvability range, indicating that many tasks are moderately susceptible to automation by generative AI tools. A significant number (16) also lie within the 50–64\% range, reflecting assignments that combine both factual and applied components. Eight assignments scored above 74\%, suggesting high susceptibility to AI completion due to the prevalence of definitional or low-complexity tasks.

\begin{figure}[hbt!]
 \centering
		\includegraphics[scale=0.4]{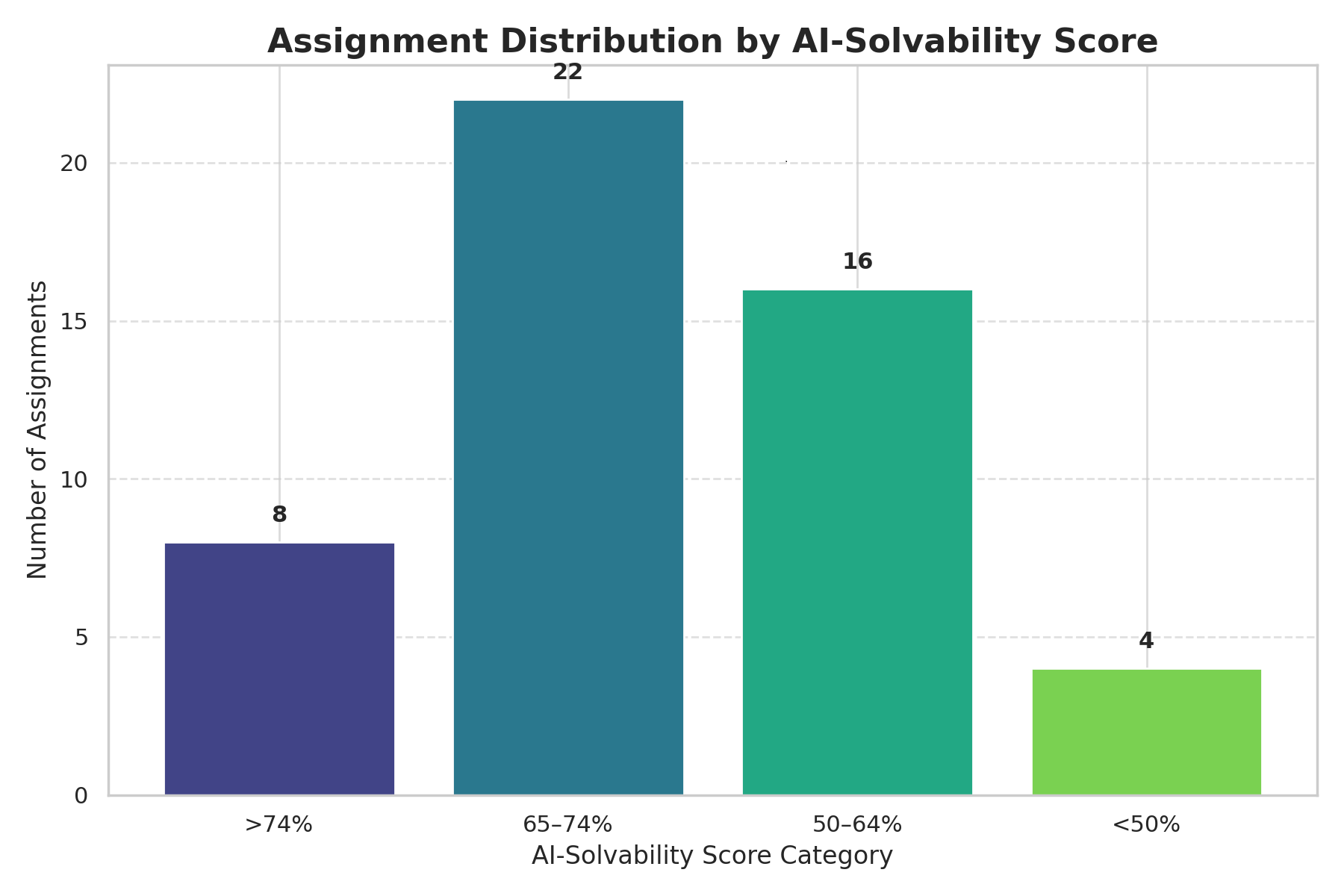}
		\caption{{Categorization of assessment categories }}
			
		\label{}
\end{figure}

In contrast, only four assignments fell below the 50\% solvability threshold, illustrating the relative scarcity of assessments requiring creative or analytical thinking that generative AI currently struggles to replicate. This distribution underscores the necessity for educators to move beyond detection-based approaches and adopt design-focused strategies that enhance the cognitive depth and originality of assessment tasks an objective directly supported by the functionality of our proposed tool.

 AI-solvability scores were computed based on the extent to which AI tools could generate accurate, relevant, and contextually appropriate responses for each task. Additionally, a categorical AI ease prediction was assigned based on the score, using defined intervals for Low (1-50\%), Medium (50–64\%), Medium-High (65–74\%), and High (\>74\%) as shown by the Table \ref{table:Categories}.

\begin{table}[hbt!]

\caption{Assignment's Categories based on AI-Solvability Score}\label{table:Categories}
\begin{tabular}{|l|l|l|}
\hline
Assignment Categories & AI-Solvability Score (\%) & AI Ease Prediction \\ \hline
1                     & \textgreater{}74\%        & High               \\ \hline
2                     & 65–74\%                   & Medium-High        \\ \hline
3                     & 50–64\%                   & Medium             \\ \hline
4                     & \textless{}50\%           & Low                \\ \hline
\end{tabular}
\end{table}

Assignments with AI-solvability scores of 70\% or higher are predominantly conceptual or definitional in nature. These tasks generally rely on established frameworks, widely accepted theories, and well-known terminology that AI models are well-trained to handle. They often lack subjective reasoning, contextual nuance, or creative design elements, making them more accessible to generative AI tools. As a result, such assignments are highly susceptible to AI-generated responses that can accurately reproduce expected content with minimal human insight. In contrast, assignments with lower AI-solvability scores, such as those around 40\%, typically involve open-ended problem-solving, real-world user scenarios, or design-based requirements. These tasks require contextual understanding, domain expertise, and creative application, making them less amenable to AI systems automation. Figures 5, 6, 7, 8, and 9 show the various sample results generated by the tool. It can be seen from those figures that the tool also provides feedback on each assignment on its strengths and weaknesses. Based on feedback, staff can improve the assessments in terms of cognitive skills before sharing them with students.

\begin{figure}[hbt!]
		\includegraphics[scale=0.6]{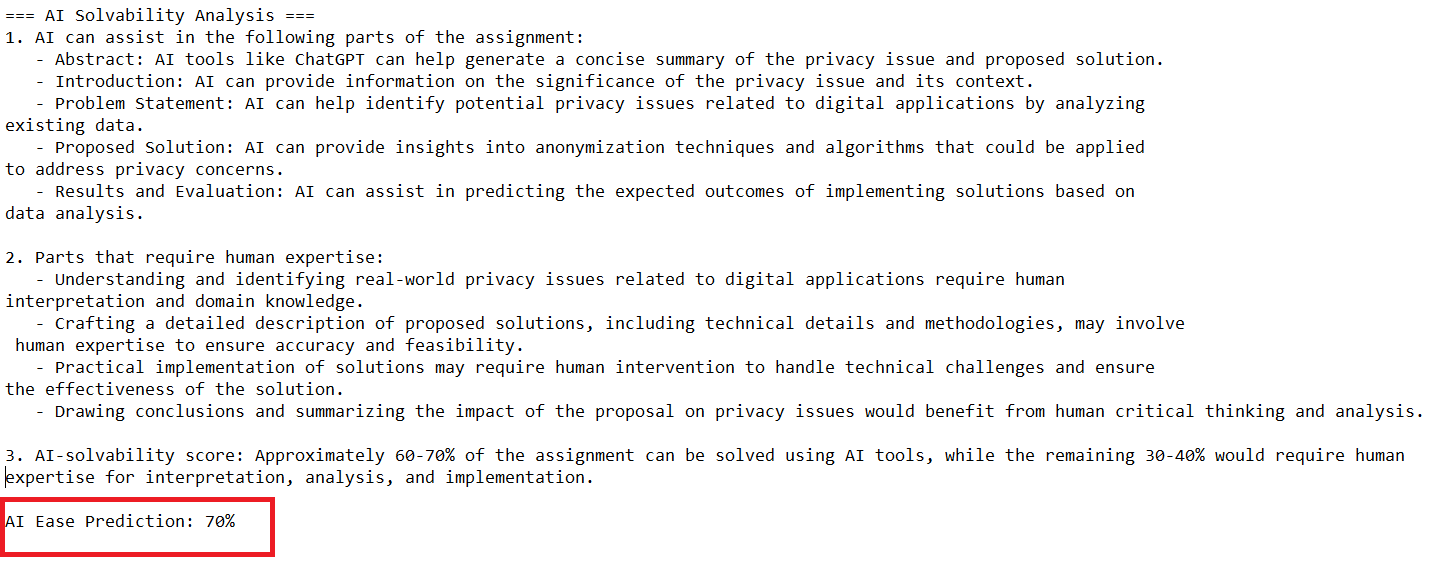}
		\caption{{Sample Result for 3 Assignments from Computer Science }}
			
		\label{}
\end{figure}

\begin{figure}[hbt!]
		\includegraphics[scale=0.6]{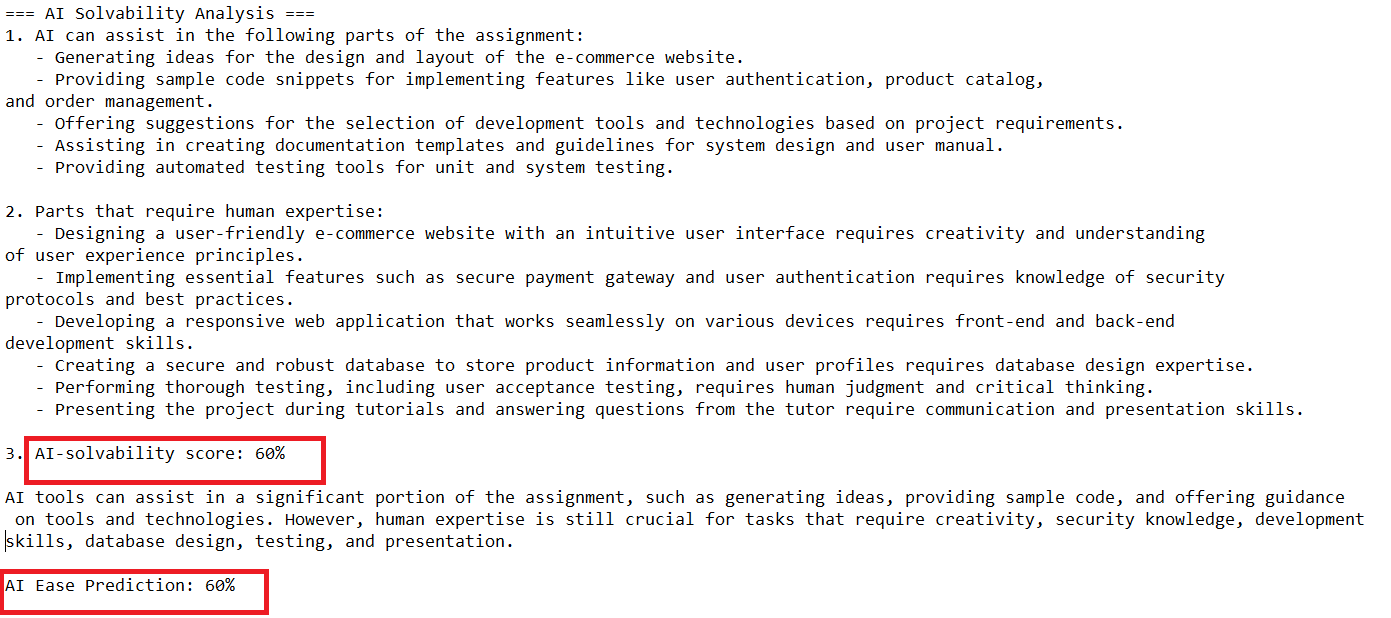}
		\caption{{Sample Result for 3 Assignments from Computer Science }}
			
		\label{}
\end{figure}

\begin{figure}[hbt!]
		\includegraphics[scale=0.6]{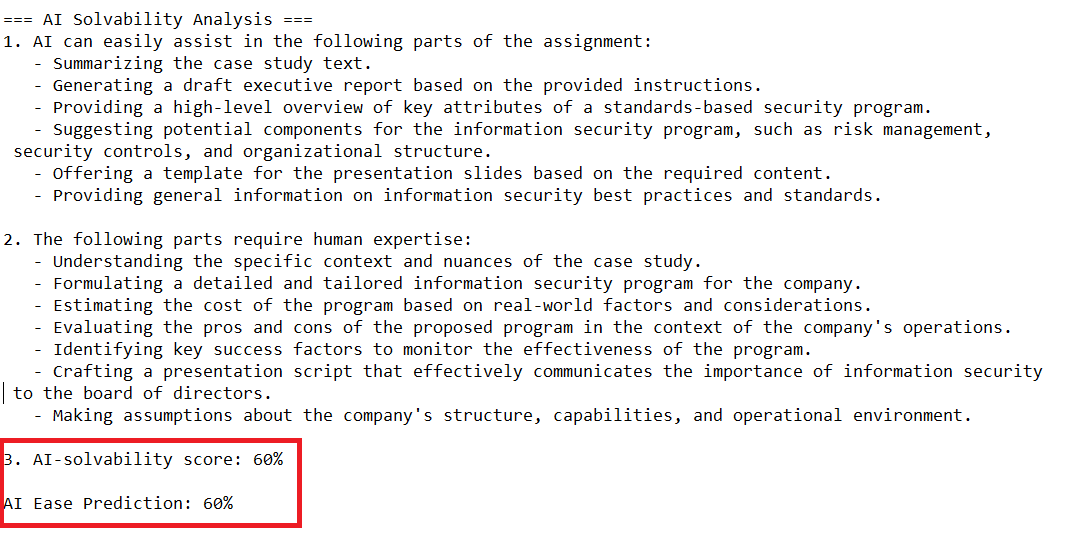}
		\caption{{Sample Results for 2 Assignments from Computer Science }}
			
		\label{}
\end{figure}

\begin{figure}[hbt!]
		\includegraphics[scale=0.6]{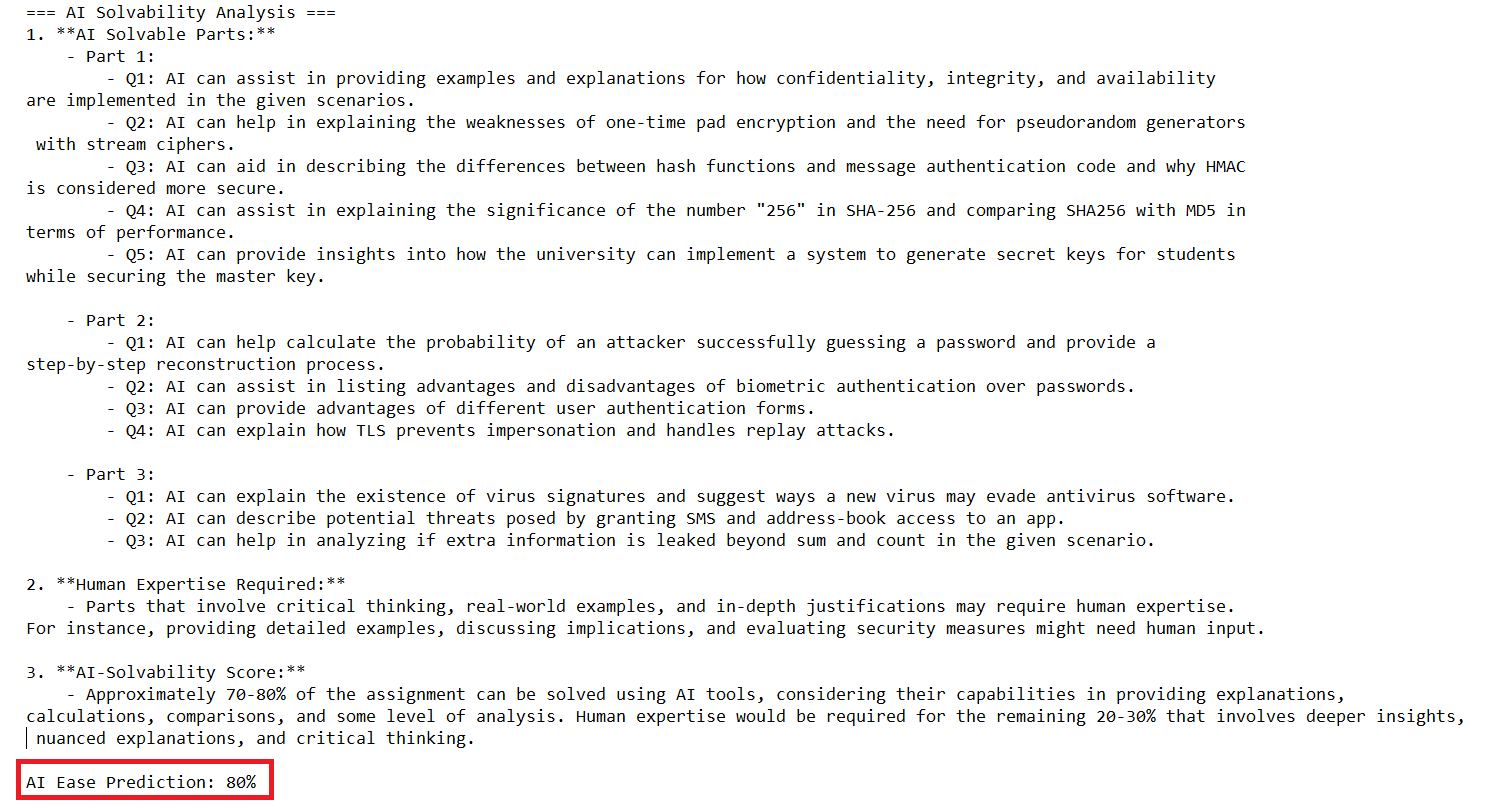}
		\caption{{Sample Results for 2 Assignments from Computer Science }}
			
		\label{}
\end{figure}

\begin{figure}[hbt!]
		\includegraphics[scale=0.6]{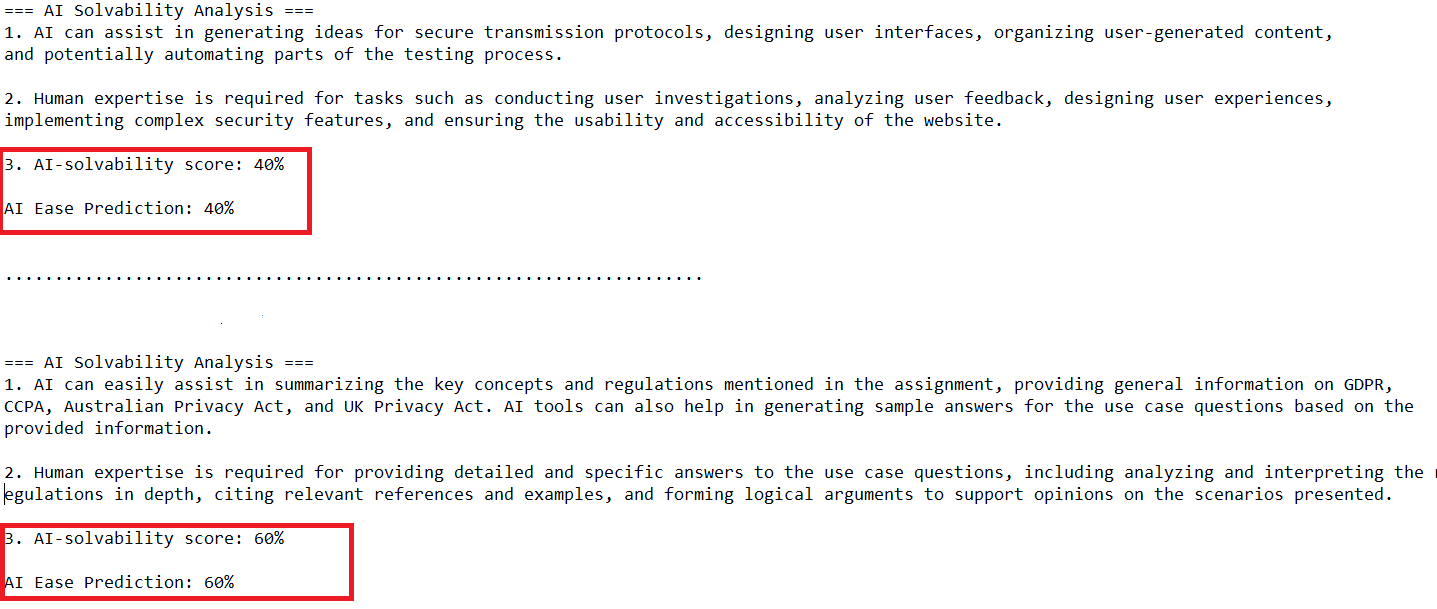}
		\caption{{Sample Results for 2 Assignments from Computer Science }}
			
		\label{}
\end{figure}


Assignments falling within the 60\% to 69\% AI-solvability range exhibit a blend of factual recall and applied reasoning. In these cases, AI can assist in generating definitions, summaries, and basic comparisons, but struggles to fully address tasks requiring system-level thinking, stakeholder considerations, or deeper analytical interpretation. While AI can support portions of these assignments, human judgment remains essential for synthesizing information and tailoring solutions to specific contexts.

This distribution highlights that AI tools are more effective at addressing lower-order cognitive tasks, particularly those aligned with the remembering and understanding levels of Bloom’s Taxonomy. However, as the cognitive complexity of assignments increases shifting toward applying, analyzing, evaluating, and creating the role of human expertise becomes increasingly critical. The findings support a tiered approach to assessment design, where integrating higher-order thinking requirements not only fosters deeper student engagement but also serves as a practical strategy for mitigating AI-dependency and preserving academic integrity.

These findings validate that our method provides a more refined
assessment of AI solvability than existing plagiarism detection tools,
making it highly effective for maintaining academic integrity.

Our approach uniquely integrates multiple AI-detection strategies, providing a far more comprehensive evaluation than traditional plagiarism checkers or AI-generated text detection tools as shown by the Table \ref{table:Comparison}. 

\begin{table}[hbt!]
\caption{Comparison of Proposed Model with Exiting}\label{table:Comparison}
\begin{tabular}{|l|l|l|l|}
\hline
\textbf{Feature}                  & \textbf{\begin{tabular}[c]{@{}l@{}}Plagiarism \\ Checkers\end{tabular}} & \textbf{\begin{tabular}[c]{@{}l@{}}GPTZero \\ (AI generated)\end{tabular}} & \textbf{Our Model} \\ \hline
Detects AI-generated text         & Yes                                                                     & Yes                                                                        & Yes                \\ \hline
Analyzes cognitive complexity     & No                                                                      & No                                                                         & Yes                \\ \hline
Integration with Bloom’s Taxonomy & No                                                                      & No                                                                         & Yes                \\ \hline
Real-time faculty feedback        & No                                                                      & Limited                                                                    & Yes                \\ \hline
Semantic similarity measurement   & No                                                                      & No                                                                         & Yes (BERT-based)   \\ \hline
AI-solvability score prediction   & No                                                                      & No                                                                         & Yes                \\ \hline
Automation and scalability        & High                                                                    & High                                                                       & High               \\ \hline
\end{tabular}
\end{table}

A key challenge of AI-assisted assessments is ensuring that students engage in original thought rather than AI-driven content generation. Our method addresses these concerns by: providing instructors with AI-solvability scores, allowing them to redesign assignments accordingly, ensuring fair AI usage, allowing AI tools to assist but not replace cognitive development, 
 minimizing false positives and negatives, ensuring students are not wrongly penalized,Encouraging responsible AI integration in education, and ensuring postgraduate students develop genuine problem-solving skills.

\subsection{Pedagogical Implications and Faculty Recommendations}
For effective implementation, we recommend:
Assignment Design Strategies: Crafting assignments that align with Bloom’s Taxonomy to minimize AI dependence, AI Detection Policies: Integrating our AI-solvability tool into Learning Management Systems (LMS) to monitor assessment integrity, faculty Training: Providing academic staff with tools and guidelines to evaluate AI-generated responses critically, adaptive AI-Resistant Questioning Techniques: Encouraging the use of multi-step reasoning and case-based learning to limit AI’s effectiveness in generating complete answers.

\section{Conclusion} 

This study responds to growing concerns over the role of generative AI in education by proposing a proactive, design-based alternative to traditional AI detection tools. Rather than relying on error-prone post-submission detection systems, our approach emphasizes the importance of designing AI-resilient assessments that promote originality, creativity, and critical thinking. Using a Python-based web tool that integrates Bloom’s Taxonomy with advanced NLP model such as GPT-3.5 Turbo, BERT-based semantic similarity, and TF-IDF educators are empowered to assess the AI-solvability of assignments in real time and receive actionable feedback for refinement.

Addressing the first research question, our findings confirm that generative AI tools have a significant impact on postgraduate students' cognitive engagement. While AI can support basic recall and explanation tasks, it risks undermining students’ development of higher-order thinking if assignments are not carefully designed. This validates concerns about the erosion of critical thinking skills in an AI-assisted academic environment.

For the second question, our analysis of of widely used AI detectitools, including commercial solutions such as Turnitin, demonstrates that these systems are fundamentally unreliable, particularly when obfuscation strategies such as paraphrasing or translation are used. Detection outputs are often vague, unverifiable, and prone to both false positives and negatives, rendering them inadequate for high-stakes academic decisions.

To answer the third research question, we successfully implemented Bloom’s Taxonomy as the foundation for our cognitive complexity framework. By evaluating whether assignments target lower-order (e.g., remembering) or higher-order (e.g., creating) cognitive processes, the tool determines their susceptibility to AI completion. This taxonomy-guided approach ensures that assessments are pedagogically aligned and resistant to automation.

Regarding the fourth question, the proposed Python-based system has proven to be a scalable and efficient solution for predicting and minimizing AI-solvability. Applied to 50 assignments from diverse computer science subdomains, the tool categorized tasks by AI vulnerability and generated solvability scores. These scores were used not only to assess assignment difficulty but also to provide real-time feedback to educators on how to revise questions for deeper learning outcomes.

Finally, in response to the fifth research question, the redesign of assessments based on cognitive complexity has been shown to enhance academic integrity significantly. Our tiered assessment analysis reveals that tasks emphasizing application, analysis, evaluation, and creation are far less likely to be solvable by AI tools. This supports a shift toward authentic assessment practices that prioritize student reasoning and innovation over AI-replicable outputs.

In summary, this research contributes a sustainable, evidence-based strategy for preserving academic integrity in the era of generative AI. By addressing assessment design rather than retroactively policing AI use, the framework fosters authentic learning, reduces reliance on unreliable detection systems, and supports the ethical integration of AI in education. Future work will explore integrating the tool into learning management systems, expanding disciplinary scope, and refining solvability models to support broader institutional adoption.


\bibliographystyle{elsarticle-num-names}
\bibliography{main}

\begin{thebibliography}{32}
\expandafter\ifx\csname natexlab\endcsname\relax\def\natexlab#1{#1}\fi
\providecommand{\url}[1]{\texttt{#1}}
\providecommand{\href}[2]{#2}
\providecommand{\path}[1]{#1}
\providecommand{\DOIprefix}{doi:}
\providecommand{\ArXivprefix}{arXiv:}
\providecommand{\URLprefix}{URL: }
\providecommand{\Pubmedprefix}{pmid:}
\providecommand{\doi}[1]{\href{http://dx.doi.org/#1}{\path{#1}}}
\providecommand{\Pubmed}[1]{\href{pmid:#1}{\path{#1}}}
\providecommand{\bibinfo}[2]{#2}
\ifx\xfnm\relax \def\xfnm[#1]{\unskip,\space#1}\fi
\bibitem[{Curzon-Hobson(2003)}]{curzon2003higher}
\bibinfo{author}{A.~Curzon-Hobson},
\newblock \bibinfo{title}{Higher learning and the critical stance},
\newblock \bibinfo{journal}{Studies in Higher Education} \bibinfo{volume}{28} (\bibinfo{year}{2003}) \bibinfo{pages}{201--212}.
\bibitem[{Garc{\'\i}a-Chitiva and Correa(2024)}]{garcia2024soft}
\bibinfo{author}{M.~d.~P. Garc{\'\i}a-Chitiva}, \bibinfo{author}{J.~C. Correa},
\newblock \bibinfo{title}{Soft skills centrality in graduate studies offerings},
\newblock \bibinfo{journal}{Studies in Higher Education} \bibinfo{volume}{49} (\bibinfo{year}{2024}) \bibinfo{pages}{956--980}.
\bibitem[{Bearman et~al.(2023)Bearman, Ryan, and Ajjawi}]{bearman2023discourses}
\bibinfo{author}{M.~Bearman}, \bibinfo{author}{J.~Ryan}, \bibinfo{author}{R.~Ajjawi},
\newblock \bibinfo{title}{Discourses of artificial intelligence in higher education: A critical literature review},
\newblock \bibinfo{journal}{Higher Education} \bibinfo{volume}{86} (\bibinfo{year}{2023}) \bibinfo{pages}{369--385}.
\bibitem[{O'Dea et~al.(2023)O'Dea, O'Dea et~al.}]{o2023artificial}
\bibinfo{author}{X.~C. O'Dea}, \bibinfo{author}{M.~O'Dea}, et~al.,
\newblock \bibinfo{title}{Is artificial intelligence really the next big thing in learning and teaching in higher education? a conceptual paper},
\newblock \bibinfo{journal}{Journal of University Teaching and Learning Practice} \bibinfo{volume}{20} (\bibinfo{year}{2023}).
\bibitem[{Essien et~al.(2024)Essien, Bukoye, O’Dea, and Kremantzis}]{essien2024influence}
\bibinfo{author}{A.~Essien}, \bibinfo{author}{O.~T. Bukoye}, \bibinfo{author}{X.~O’Dea}, \bibinfo{author}{M.~Kremantzis},
\newblock \bibinfo{title}{The influence of ai text generators on critical thinking skills in uk business schools},
\newblock \bibinfo{journal}{Studies in Higher Education} \bibinfo{volume}{49} (\bibinfo{year}{2024}) \bibinfo{pages}{865--882}.
\bibitem[{Boutyour et~al.(2024)Boutyour, Idrissi, and Uden}]{boutyour2024artificial}
\bibinfo{author}{Y.~Boutyour}, \bibinfo{author}{A.~Idrissi}, \bibinfo{author}{L.~Uden},
\newblock \bibinfo{title}{Artificial intelligence and assessment generators in education: A comprehensive review},
\newblock \bibinfo{journal}{Modern Artificial Intelligence and Data Science 2024: Tools, Techniques and Systems}  (\bibinfo{year}{2024}) \bibinfo{pages}{265--284}.
\bibitem[{Rudolph et~al.(2023)Rudolph, Tan, and Tan}]{rudolph2023chatgpt}
\bibinfo{author}{J.~Rudolph}, \bibinfo{author}{S.~Tan}, \bibinfo{author}{S.~Tan},
\newblock \bibinfo{title}{Chatgpt: Bullshit spewer or the end of traditional assessments in higher education?},
\newblock \bibinfo{journal}{Journal of applied learning and teaching} \bibinfo{volume}{6} (\bibinfo{year}{2023}) \bibinfo{pages}{342--363}.
\bibitem[{Charles et~al.(2025)Charles, Yousuf, Chua, Matthews, Harnett, and Hinton}]{charles2025ai}
\bibinfo{author}{K.~A. Charles}, \bibinfo{author}{A.~Yousuf}, \bibinfo{author}{H.~C. Chua}, \bibinfo{author}{S.~Matthews}, \bibinfo{author}{J.~Harnett}, \bibinfo{author}{T.~Hinton},
\newblock \bibinfo{title}{Ai in action: Changes to student perceptions when using generative artificial intelligence for the creation of a multimedia project-based assessment.},
\newblock \bibinfo{journal}{European Journal of Pharmacology}  (\bibinfo{year}{2025}) \bibinfo{pages}{177508}.
\bibitem[{Pliuskuvien{\.e} et~al.(2024)Pliuskuvien{\.e}, Radvilait{\.e}, Juodagalvyt{\.e}, Ramanauskait{\.e}, and Stefanovi{\v{c}}}]{pliuskuviene2024educational}
\bibinfo{author}{B.~Pliuskuvien{\.e}}, \bibinfo{author}{U.~Radvilait{\.e}}, \bibinfo{author}{R.~Juodagalvyt{\.e}}, \bibinfo{author}{S.~Ramanauskait{\.e}}, \bibinfo{author}{P.~Stefanovi{\v{c}}},
\newblock \bibinfo{title}{Educational data mining and learning analytics: Text generators usage effect on students’ grades},
\newblock \bibinfo{journal}{New Trends in Computer Sciences} \bibinfo{volume}{2} (\bibinfo{year}{2024}) \bibinfo{pages}{19--30}.
\bibitem[{Yitages and Kasai(2024)}]{yitages2024faculty}
\bibinfo{author}{M.~Yitages}, \bibinfo{author}{A.~Kasai},
\newblock \bibinfo{title}{Faculty opinions of ai tools: Text generators and machine translators.},
\newblock \bibinfo{journal}{American Journal of Undergraduate Research} \bibinfo{volume}{20} (\bibinfo{year}{2024}).
\bibitem[{Gooch et~al.(2024)Gooch, Waugh, Richards, Slaymaker, and Woodthorpe}]{gooch2024exploring}
\bibinfo{author}{D.~Gooch}, \bibinfo{author}{K.~Waugh}, \bibinfo{author}{M.~Richards}, \bibinfo{author}{M.~Slaymaker}, \bibinfo{author}{J.~Woodthorpe},
\newblock \bibinfo{title}{Exploring the profile of university assessments flagged as containing ai-generated material},
\newblock \bibinfo{journal}{ACM Inroads} \bibinfo{volume}{15} (\bibinfo{year}{2024}) \bibinfo{pages}{39--47}.
\bibitem[{Vetter et~al.(2024)Vetter, Lucia, Jiang, and Othman}]{vetter2024towards}
\bibinfo{author}{M.~A. Vetter}, \bibinfo{author}{B.~Lucia}, \bibinfo{author}{J.~Jiang}, \bibinfo{author}{M.~Othman},
\newblock \bibinfo{title}{Towards a framework for local interrogation of ai ethics: A case study on text generators, academic integrity, and composing with chatgpt},
\newblock \bibinfo{journal}{Computers and composition} \bibinfo{volume}{71} (\bibinfo{year}{2024}) \bibinfo{pages}{102831}.
\bibitem[{Gamage et~al.(2023)Gamage, Dehideniya, Xu, and Tang}]{gamage2023chatgpt}
\bibinfo{author}{K.~A. Gamage}, \bibinfo{author}{S.~C. Dehideniya}, \bibinfo{author}{Z.~Xu}, \bibinfo{author}{X.~Tang},
\newblock \bibinfo{title}{Chatgpt and higher education assessments: More opportunities than concerns?},
\newblock \bibinfo{journal}{Journal of Applied Learning and Teaching} \bibinfo{volume}{6} (\bibinfo{year}{2023}) \bibinfo{pages}{358--369}.
\bibitem[{Banda et~al.(2023)Banda, Phiri, Kaale, Banda, Mpolomoka, Chikopela, and Mushibwe}]{banda2023application}
\bibinfo{author}{S.~Banda}, \bibinfo{author}{F.~Phiri}, \bibinfo{author}{J.~Kaale}, \bibinfo{author}{A.~M. Banda}, \bibinfo{author}{D.~Mpolomoka}, \bibinfo{author}{R.~Chikopela}, \bibinfo{author}{C.~Mushibwe},
\newblock \bibinfo{title}{Application of bloom’s taxonomy in categorization of cognitive process development in colleges},
\newblock \bibinfo{journal}{Journal of Education and Practice} \bibinfo{volume}{14} (\bibinfo{year}{2023}) \bibinfo{pages}{6--13}.
\bibitem[{Weber-Wulff et~al.(2023)Weber-Wulff, Anohina-Naumeca, Bjelobaba, Folt{\`y}nek, Guerrero-Dib, Popoola, {\v{S}}igut, and Waddington}]{weber2023testing}
\bibinfo{author}{D.~Weber-Wulff}, \bibinfo{author}{A.~Anohina-Naumeca}, \bibinfo{author}{S.~Bjelobaba}, \bibinfo{author}{T.~Folt{\`y}nek}, \bibinfo{author}{J.~Guerrero-Dib}, \bibinfo{author}{O.~Popoola}, \bibinfo{author}{P.~{\v{S}}igut}, \bibinfo{author}{L.~Waddington},
\newblock \bibinfo{title}{Testing of detection tools for ai-generated text},
\newblock \bibinfo{journal}{International Journal for Educational Integrity} \bibinfo{volume}{19} (\bibinfo{year}{2023}) \bibinfo{pages}{1--39}.
\bibitem[{Gao et~al.(2022)Gao, Howard, Markov, Dyer, Ramesh, Luo, and Pearson}]{gao2022comparing}
\bibinfo{author}{C.~A. Gao}, \bibinfo{author}{F.~M. Howard}, \bibinfo{author}{N.~S. Markov}, \bibinfo{author}{E.~C. Dyer}, \bibinfo{author}{S.~Ramesh}, \bibinfo{author}{Y.~Luo}, \bibinfo{author}{A.~T. Pearson},
\newblock \bibinfo{title}{Comparing scientific abstracts generated by chatgpt to original abstracts using an artificial intelligence output detector, plagiarism detector, and blinded human reviewers},
\newblock \bibinfo{journal}{BioRxiv}  (\bibinfo{year}{2022}) \bibinfo{pages}{2022--12}.
\bibitem[{Anderson et~al.(2023)Anderson, Belavy, Perle, Hendricks, Hespanhol, Verhagen, and Memon}]{anderson2023ai}
\bibinfo{author}{N.~Anderson}, \bibinfo{author}{D.~L. Belavy}, \bibinfo{author}{S.~M. Perle}, \bibinfo{author}{S.~Hendricks}, \bibinfo{author}{L.~Hespanhol}, \bibinfo{author}{E.~Verhagen}, \bibinfo{author}{A.~R. Memon}, \bibinfo{title}{Ai did not write this manuscript, or did it? can we trick the ai text detector into generated texts? the potential future of chatgpt and ai in sports \& exercise medicine manuscript generation}, \bibinfo{year}{2023}.
\bibitem[{Elkhatat et~al.(2023)Elkhatat, Elsaid, and Almeer}]{elkhatat2023evaluating}
\bibinfo{author}{A.~M. Elkhatat}, \bibinfo{author}{K.~Elsaid}, \bibinfo{author}{S.~Almeer},
\newblock \bibinfo{title}{Evaluating the efficacy of ai content detection tools in differentiating between human and ai-generated text},
\newblock \bibinfo{journal}{International Journal for Educational Integrity} \bibinfo{volume}{19} (\bibinfo{year}{2023}) \bibinfo{pages}{17}.
\bibitem[{Demers(16)}]{demers16best}
\bibinfo{author}{T.~Demers}, \bibinfo{title}{of the best ai and chatgpt content detectors compared. search engine land.(2023, april 25)}, \bibinfo{year}{16}.
\bibitem[{Gewirtz(2023)}]{gewirtz2023can}
\bibinfo{author}{D.~Gewirtz}, \bibinfo{title}{Can ai detectors save us from chatgpt? i tried 5 online tools to find out. zdnet}, \bibinfo{year}{2023}.
\bibitem[{Krishna et~al.(2023)Krishna, Song, Karpinska, Wieting, and Iyyer}]{krishna2023paraphrasing}
\bibinfo{author}{K.~Krishna}, \bibinfo{author}{Y.~Song}, \bibinfo{author}{M.~Karpinska}, \bibinfo{author}{J.~Wieting}, \bibinfo{author}{M.~Iyyer},
\newblock \bibinfo{title}{Paraphrasing evades detectors of ai-generated text, but retrieval is an effective defense},
\newblock \bibinfo{journal}{Advances in Neural Information Processing Systems} \bibinfo{volume}{36} (\bibinfo{year}{2023}) \bibinfo{pages}{27469--27500}.
\bibitem[{Pegoraro et~al.(2023)Pegoraro, Kumari, Fereidooni, and Sadeghi}]{pegoraro2023chatgpt}
\bibinfo{author}{A.~Pegoraro}, \bibinfo{author}{K.~Kumari}, \bibinfo{author}{H.~Fereidooni}, \bibinfo{author}{A.-R. Sadeghi},
\newblock \bibinfo{title}{To chatgpt, or not to chatgpt: That is the question!},
\newblock \bibinfo{journal}{arXiv preprint arXiv:2304.01487}  (\bibinfo{year}{2023}).
\bibitem[{Walters(2023)}]{walters2023effectiveness}
\bibinfo{author}{W.~H. Walters},
\newblock \bibinfo{title}{The effectiveness of software designed to detect ai-generated writing: A comparison of 16 ai text detectors},
\newblock \bibinfo{journal}{Open Information Science} \bibinfo{volume}{7} (\bibinfo{year}{2023}) \bibinfo{pages}{20220158}.
\bibitem[{{\v{S}}IGUT(????)}]{vsigutevaluation}
\bibinfo{author}{P.~{\v{S}}IGUT},
\newblock \bibinfo{title}{Evaluation of machine-generated text detectors}  (????).
\bibitem[{Gagliardi(2024)}]{gagliardi2024natural}
\bibinfo{author}{G.~Gagliardi},
\newblock \bibinfo{title}{Natural language processing techniques for studying language in pathological ageing: A scoping review},
\newblock \bibinfo{journal}{International Journal of Language \& Communication Disorders} \bibinfo{volume}{59} (\bibinfo{year}{2024}) \bibinfo{pages}{110--122}.
\bibitem[{Carrasco and Dias(2024)}]{carrasco2024enhancing}
\bibinfo{author}{P.~Carrasco}, \bibinfo{author}{S.~Dias},
\newblock \bibinfo{title}{Enhancing restaurant management through aspect-based sentiment analysis and nlp techniques},
\newblock \bibinfo{journal}{Procedia Computer Science} \bibinfo{volume}{237} (\bibinfo{year}{2024}) \bibinfo{pages}{129--137}.
\bibitem[{Gue et~al.(2024)Gue, Rahim, Rojas-Carabali, Agrawal, Rk, Abisheganaden, and Yip}]{gue2024evaluating}
\bibinfo{author}{C.~C.~Y. Gue}, \bibinfo{author}{N.~D.~A. Rahim}, \bibinfo{author}{W.~Rojas-Carabali}, \bibinfo{author}{R.~Agrawal}, \bibinfo{author}{P.~Rk}, \bibinfo{author}{J.~Abisheganaden}, \bibinfo{author}{W.~F. Yip},
\newblock \bibinfo{title}{Evaluating the openai’s gpt-3.5 turbo’s performance in extracting information from scientific articles on diabetic retinopathy},
\newblock \bibinfo{journal}{Systematic reviews} \bibinfo{volume}{13} (\bibinfo{year}{2024}) \bibinfo{pages}{135}.
\bibitem[{Wang et~al.(2024)Wang, Wang, Huang, Liu, Yang, Liao, Lu, and Wu}]{wang2024investigation}
\bibinfo{author}{W.~h. Wang}, \bibinfo{author}{S.~y. Wang}, \bibinfo{author}{J.~Y. Huang}, \bibinfo{author}{X.~d. Liu}, \bibinfo{author}{J.~Yang}, \bibinfo{author}{M.~Liao}, \bibinfo{author}{Q.~Lu}, \bibinfo{author}{Z.~Wu},
\newblock \bibinfo{title}{An investigation study on the interpretation of ultrasonic medical reports using openai's gpt-3.5-turbo model},
\newblock \bibinfo{journal}{Journal of Clinical Ultrasound} \bibinfo{volume}{52} (\bibinfo{year}{2024}) \bibinfo{pages}{105--111}.
\bibitem[{Zunlan and Xiaohong(2025)}]{zunlan2025dynamic}
\bibinfo{author}{X.~Zunlan}, \bibinfo{author}{N.~Xiaohong},
\newblock \bibinfo{title}{Dynamic bert-svm hybrid model for enhanced semantic similarity evaluation in english teaching texts},
\newblock \bibinfo{journal}{Journal of English Language Teaching and Applied Linguistics} \bibinfo{volume}{7} (\bibinfo{year}{2025}) \bibinfo{pages}{01--11}.
\bibitem[{Chen(2025)}]{chen2025semantic}
\bibinfo{author}{S.~Chen},
\newblock \bibinfo{title}{Semantic relationship extraction of english long sentences and quality optimization of machine translation based on bert model},
\newblock \bibinfo{journal}{Journal of Computational Methods in Sciences and Engineering}  (\bibinfo{year}{2025}) \bibinfo{pages}{14727978251322656}.
\bibitem[{Lu(2024)}]{lu2024enhancing}
\bibinfo{author}{J.~Lu},
\newblock \bibinfo{title}{Enhancing chatbot user satisfaction: A machine learning approach integrating decision tree, tf-idf, and bertopic},
\newblock in: \bibinfo{booktitle}{2024 IEEE 6th International Conference on Power, Intelligent Computing and Systems (ICPICS)}, \bibinfo{organization}{IEEE}, \bibinfo{year}{2024}, pp. \bibinfo{pages}{823--828}.
\bibitem[{Al~Tawil et~al.(2024)Al~Tawil, Almazaydeh, Qawasmeh, Qawasmeh, Alshinwan, and Elleithy}]{al2024comparative}
\bibinfo{author}{A.~Al~Tawil}, \bibinfo{author}{L.~Almazaydeh}, \bibinfo{author}{D.~Qawasmeh}, \bibinfo{author}{B.~Qawasmeh}, \bibinfo{author}{M.~Alshinwan}, \bibinfo{author}{K.~Elleithy},
\newblock \bibinfo{title}{Comparative analysis of machine learning algorithms for email phishing detection using tf-idf, word2vec, and bert},
\newblock \bibinfo{journal}{Comput. Mater. Contin} \bibinfo{volume}{81} (\bibinfo{year}{2024}) \bibinfo{pages}{3395}.

\end{thebibliography}

\end{document}